\documentstyle[prd,aps]{revtex}
\begin{document}
\draft
%
%
\twocolumn[\hsize\textwidth\columnwidth\hsize\csname
@twocolumnfalse\endcsname

\preprint{SUSSEX-AST 96/8-2, RCG-96/10, astro-ph/9608106}
\title{Normalization of modes in an open universe}
\author{Juan Garc\'{\i}a-Bellido}
\address{Astronomy Centre, University of Sussex, Falmer, 
Brighton BN1 9QH, United Kingdom}
\author{Andrew R. Liddle}
\address{Astronomy Centre, University of Sussex, Falmer, 
Brighton BN1 9QH, United Kingdom} 
\author{David H. Lyth}
\address{School of Physics and Materials, University of Lancaster, 
Lancaster LA1 4YB, United Kingdom}
\author{David Wands}
\address{School of Mathematical Studies, University of Portsmouth,
Portsmouth PO1 2EG, United Kingdom}
\date{\today}
\maketitle
\begin{abstract}
We discuss the appropriate normalization of modes required to generate
a homogeneous random field in an open Friedmann-Robertson-Walker
universe. We consider scalar random fields and certain tensor random
fields that can be obtained by covariantly differentiating a scalar.
Modes of interest fall into three categories: the familiar sub-curvature 
modes, the more recently discussed super-curvature modes, and a set of 
discrete modes with positive eigenvalues which can be used to generate 
homogeneous tensor random fields even though the underlying scalar field is 
not homogeneous.  A particular example of the last case which has been 
discussed in the literature is the bubble wall fluctuation in open 
inflationary universes.
\end{abstract}
\pacs{PACS numbers: 98.80.Cq \hspace*{1cm} Preprint SUSSEX-AST 96/8-2, 
RCG-96/10, astro-ph/9608106}
 
\vskip2pc]

\section{Introduction}

Recently, considerable attention has been directed to models of
structure formation in an open universe~\cite{openlss}. The description of 
perturbations in an open universe is more subtle than in the spatially
flat case, because one has to make a distinction between the concept of
a {\em function} and that of a {\em random field}. The latter can be
thought of as an ensemble of functions with a probability assigned to
each one.  In cosmology we are interested in some finite region around
us, possibly much bigger than the Hubble distance. A given
perturbation about the homogeneous background, say the density
perturbation, is some function of space-time in this region, but the
specific form of the function is not thought to be very important.
Rather, we make the hypothesis that it is a typical realization of a
random field. It is assumed that the field is {\em homogeneous} with
respect to the translations and rotations of the coordinate system. It
is usually also assumed to be {\em Gaussian}, which means that there
exists an expansion into mode functions with independent Gaussian
probability distributions for the coefficients. {}From now on we usually
take the homogeneity and Gaussianity as read, referring simply to a
`random field'.

Let us first recall the situation for a scalar field, such as the
density perturbation. In order to generate a random field (and also to
decouple the time dependence of the modes in the linear
approximation) the mode functions will be eigenfunctions of the
Laplacian in comoving coordinates.  In a flat universe, the most general
square-integrable {\em function} may be expanded in terms of the
eigenfunctions with $k^2>0$ (as usual $-k^2$ is the eigenvalue of the
Laplacian). In that case these same functions also give the most general
scalar random field. In an open universe the situation is different. In
units of the curvature scale, the most general square-integrable
function may be expanded using only the $k^2>1$ modes (sub-curvature
modes). In particular a perturbation defined in any finite region can be
so expanded, which is all we need for any cosmological application.  But
now the situation for a random field is different; to obtain the most
general scalar random field one needs, in addition, the $0< k^2 \leq1$
modes (super-curvature modes). This fact has been known to
mathematicians for half a century \cite{krein,yaglom}, though it has
only recently been brought to the attention of the astrophysics
community~\cite{LW}. One of the objects of the present paper is to
derive, in a simple way, the normalization of the mode functions which
will lead to a homogeneous random field (with a natural choice for the
variances of the coefficients).

In cosmology we might also be interested in vector and (second rank or
higher) tensor functions, and the corresponding random fields.  In
general such objects have to be expanded in terms of different mode
functions, which we shall not consider. But an important special case
arises when the vector or tensor is the spatial derivative of a scalar.
The central purpose of this paper is to treat this case, and focus on a
fact concerning the open universe which does not seem to have been
discussed in the mathematics literature.  Namely, that a homogeneous
vector or tensor random field can be constructed by differentiating an
{\em inhomogeneous} scalar random field. We conjecture that the most
general random field derived from a scalar can be expanded in
terms of the continuum modes plus some new mode functions with
$k^2=1-n^2$ (discrete modes) where $n$ is an integer. We support our
conjecture by displaying an explicit normalization of the mode functions
for the cases $k^2=0$ and $k^2=-3$ which is suitable for generating a
tensor random field. Examples of both cases exist in the recent
literature. A discrete mode with $k^2=0$ is generated by fluctuations
a massless scalar field in de Sitter space-time~\cite{sasaki} and the
effect on the microwave background of such a mode was evaluated in
Ref.~\cite{GZ}. A discrete $k^2=-3$ mode can be generated by
fluctuations of the bubble wall~\cite{Hamazaki,bubble}, in single-bubble
models of open universe inflation~\cite{open}. 

To be more specific, we write down the mode expansion for a scalar
random field in terms of eigenfunctions of the spatial Laplacian,
\mbox{$\nabla^2 Z_{klm}=-k^2 Z_{klm}$}: 
\begin{equation} 
\label{modeexp}
f(r,\theta,\phi)= \int dk \sum_{l=0}^\infty \sum_{m=-l}^l
        f_{klm} Z_{klm}(r,\theta,\phi) \, ,
\end{equation}
where the $f_{klm}$ are members of an ensemble.  The possible values
of $k$ will be discussed in what follows.  We will consider both a
continuous spectrum of modes and discrete values of $k$.  We work in a
spherical coordinate system with line element
\begin{equation}
\label{lineelement}
ds^2= dr^2 + \sinh^2r(d\theta^2+\sin^2\theta d\phi^2) \, ,
\end{equation}
corresponding to the homogeneous spatial hypersurfaces in an open
Friedmann--Robertson--Walker universe. The normalized eigenfunctions
can be written as
\begin{equation}
\label{Zklm}
Z_{klm}({\bf x}) = \Pi_{kl}(r) Y_{lm}(\theta,\phi) \,,
\end{equation}
where $Y_{lm}(\theta,\phi)$ are the usual spherical harmonics on the
two-sphere \cite{PTVF} and $\Pi_{kl}(r)$ are eigenfunctions of the operator
\begin{equation}
\frac{1}{\sinh^2r} \, \frac{d}{dr} \, \left(\sinh^2 r \, \frac{d}{dr} \,
        \right) + \frac{l(l+1)}{\sinh^2 r} \;.
\end{equation}
The normalization of the $\Pi_{kl}(r)$ is the subject of the present paper.

In a Gaussian random field, if one uses the complex form of the spherical 
harmonics the magnitudes $|f_{klm}|$ of each coefficient have independent 
Gaussian probability distributions for $m \geq 0$. The reality condition
\begin{equation}
f_{kl,-m} \Pi_{kl} =(-1)^m \, \left( f_{klm} \Pi_{kl} \right)^* \,,
\end{equation}
which follows from $Y_{l,-m}(\theta,\phi) = (-1)^m
Y^*_{lm}(\theta,\phi)$, fixes the phase of the $m = 0$ modes, while
the other modes have uniformly distributed random phases subject to
the above equation. In what follows we will find it more convenient to
use the real form of the $Y_{lm}(\theta,\phi)$; the coefficients then
have fixed phases, $\arg(f_{klm}) = -\arg(\Pi_{kl})$, and the
magnitudes of $f_{klm}$ are independent random variables for all
$m$ from $-l$ to $+l$.

The variance of the distribution is defined by
\begin{equation}
\label{indep}
\langle f^*_{klm} f_{k'l'm'} \rangle = A_k {\cal P}(k) \delta(k-k')
        \delta_{ll'} \delta_{mm'} \,.
\end{equation}
where the brackets denote an ensemble average. As we shall discuss in the 
next subsection, the variance is taken to be independent of $l$ and
$m$ to allow us to construct a homogeneous random field. Several different 
conventions exist in the literature concerning the definition of the power 
spectrum ${\cal P}(k)$, corresponding to different choices of the prefactor 
$A_k$~\cite{LW}. In what follows we shall not need to define a particular 
separation. Since the field is Gaussian, it is completely defined by the 
two-point correlation function. For the continuous case this is defined by
\begin{equation}
\label{corrfunc}
\langle f^*({\bf x}_1) f({\bf x}_2) \rangle = \int dk \,A_k\,{\cal P}(k)\,
        \sum_{lm} \, Z^*_{klm}({\bf x}_1)\,Z_{klm}({\bf x}_2) \,,
\end{equation}
and in the discrete case the integral over $k$ is replaced by a
sum. Here and throughout we use the abbreviated notation
\begin{equation}
\sum_{lm} \equiv \sum_{l=0}^{\infty} \, \sum_{m=-l}^l \,.
\end{equation}

\section{Unitarity and homogeneity}

The above procedure generates a scalar Gaussian random field.  The field is
said to be homogeneous if the correlation function depends only on the
geodesic distance between the two points. By covariantly differentiating 
such a field, one can obtain a tensor random field which is likewise 
homogeneous if its correlation function is unaffected by a coordinate change 
(except of course for the transformation that the change induces in the 
components of the tensor). It is obvious that differentiating a homogeneous 
scalar random field always gives a homogeneous tensor random field, but we
shall demonstrate that in some circumstances one can also obtain a
homogeneous tensor field by differentiating an {\em inhomogeneous}
scalar field.

The central purpose of this paper is to ask what restriction is placed
on the normalization of the modes by the requirement that the random
field be homogeneous, and to discuss which values of $k$ are
compatible with this requirement. The homogeneity requirement, that
the correlation function depends only on the geodesic distance between
the points, is equivalent to the requirement that under a shift in the
origin or orientation of the coordinate system the Gaussianity is
preserved and the correlation function is unaltered. We begin by
showing that for a scalar field this is the case if and only if
such a coordinate change corresponds to a unitary transformation of
the basis functions.\footnote{The invariance of the correlation
function was shown in Ref.~\cite{LW}. As in that reference we treat
infinite sums as finite, which should be valid if the infinite sum
Eq.~(\ref{corrfunc}) is uniformly convergent. That has been
demonstrated in Ref.~\cite{sasaki} for $k^2>0$, but we have not
investigated the question for $k^2\leq 0$.}

Since a coordinate change does not affect the eigenvalues of the Laplacian, 
it will be useful to consider a field constructed from modes with a single 
value of $k$
\begin{equation}
\label{discretemodeexp}
f(r,\theta,\phi) = \sum_{lm} f_{klm} Z_{klm} (r,\theta,\phi) \,.
\end{equation}
The joint probability distribution for the coefficients is
\begin{equation}
\label{jointprob}
{\rm probability} = N \exp \left(- \,
        \frac{\sum_{lm}|f_{klm}|^2}{2A_k{\cal P}(k)} \right) \,,
\end{equation}
(where the normalization factor $N$ is actually infinitesimal because the
sum is infinite).

Under a change of coordinates, the mode functions undergo a linear 
transformation
\begin{equation}
\label{Zunitary}
Z_{klm}(r,\theta,\phi) = \sum_{l'm'} U^k_{lml'm'}
        Z_{kl'm'}(r',\theta',\phi')\,,
\end{equation}
and the mode expansion in Eq.~(\ref{discretemodeexp}) becomes
\begin{eqnarray}
f(r,\theta,\phi) & = &  \sum_{lm} f_{klm} \, \sum_{l'm'}
        U^k_{lml'm'} Z_{kl'm'}(r',\theta',\phi') \nonumber \\
& = & \sum_{l'm'} f'_{kl'm'} Z_{kl'm'}(r',\theta',\phi') \, ,
\end{eqnarray}
where
\begin{equation}
\label{fprime}
f'_{kl'm'} = \sum_{lm} U^k_{lml'm'} f_{klm}\,.
\label{ftrans} 
\end{equation}

The form of the joint probability distribution, Eq.~(\ref{jointprob}),
with respect to the transformed coefficients is clearly unaltered if
and only if
\begin{equation}
\sum_{l'm'}|f'_{kl'm'}|^2 = \sum_{lm} |f_{klm}|^2 \,,
\end{equation}
which requires the matrix $U$ to be unitary:
\begin{equation}
\label{unitary}
\sum_{lm} (U^k_{lml''m''})^*\,U^k_{lml'm'}
        = \delta_{l'l''} \delta_{m'm''}\, .
\end{equation}
Although we have shown this only for modes with a single value of $k$,
the corresponding expression for a continuous spectrum of modes will
also be unaffected, since the transformation does not act on $k$. One
can show that the correlation function defined by
Eq.~(\ref{corrfunc}), or the spectrum in Eq.~(\ref{indep}), are also
unaffected for a unitary transformation between modes.  Thus the field
is homogeneous if and only if $U$ is unitary.

The unitarity requirement allows one to change the normalization of
the modes by an arbitrary $k$-dependent real factor, and a completely
arbitrary phase. What is important is the dependence of the {\em
magnitude} of the normalization factor on $l$ and $m$. The normalization of 
the spherical harmonics $Y_{lm}(\theta,\phi)$ ensures that they transform 
unitarily under an arbitrary rotation (and hence the distribution defined in 
Eq.~(\ref{indep}) is isotropic), but we have to ensure that the complete 
basis functions $Z_{klm}$ transform unitarily under both rotations and 
shifts of origin.

Note that a rotation about a fixed origin leaves $k$ and $l$ fixed but
mixes the different $m$-multipoles, while a shift along the
$\theta=0$ axis leaves $k$ and $m$ fixed but mixes different $l$
multipoles. Because an arbitrary shift and rotation of the
origin can be decomposed into a rotation, followed by a shift along
$\theta=0$, followed by another rotation, the unitarity of the
spherical harmonics under rotations fixes the $m$-dependence of the
normalization. We now seek the correct $l$-dependence of the
normalization of the radial functions $\Pi_{kl}(r)$ which ensures
homogeneity under a shift of the origin.

\section{Homogeneous scalar random fields}

\subsection*{Sub-curvature modes}

For modes with $k^2 > 1$, we can split $\Pi_{kl}(r)$ as
\begin{equation}
\Pi_{kl}(r) = N_{kl} \, \tilde\Pi_{kl}(r)  \,,
\end{equation}
where $N_{kl}$ is a normalization factor, to be determined, and 
$\tilde\Pi_{kl}(r)$ are the unnormalized functions \cite{Har}
\begin{equation}
\label{Pikl}
\tilde\Pi_{kl}(r) = q^{-2} (\sinh r)^l \left(\frac{-1}{\sinh r} 
        \frac{d}{dr} \right)^{l+1} \cos(qr) \, ,
\end{equation}
where $q^2 = k^2-1$. For these modes, $q=\pm\sqrt{k^2-1}$ is real and
the eigenfunctions are exponentially decreasing beyond the curvature
scale [which equals unity for the line element in Eq.~(\ref{lineelement})].  
It is possible to choose the normalization factor $N_{kl}$ to give an 
orthogonality relation between the modes (not necessarily
orthonormality) of the form
\begin{equation}
\label{norm}
\int dV  Z^*_{klm}({\bf x}) Z_{k'l'm'}({\bf x}) = B_k \,\delta(|q|-|q'|)\,
        \delta_{ll'}\,\delta_{mm'} \, ,
\end{equation}
where $B_k$ is a finite real function of $k$, independent of $l$ and
$m$. This relation ensures unitarity of the matrix $U$ for a coordinate 
shift~\cite{LW}, as we show in Appendix~A.  In particular, the usual 
normalization factor is taken to be~\cite{Har}
\begin{equation}
\label{subNkl}
N_{kl} = \sqrt\frac2\pi \, q^2 \, \prod^l_{s=0}(s^2+q^2)^{-1/2}\, ,
\end{equation}
corresponding to $B_k=1$ in Eq.~(\ref{norm}). 

Note that the radial functions $\tilde\Pi_{kl}(r)$ and the normalization 
Eq.~(\ref{subNkl}) are real, and one can always use the real form of the 
spherical harmonics. With that convention, $U$ is real and unitarity
of the transformation, given in Eq.~(\ref{unitary}), implies that the
matrix is also orthogonal;
\begin{equation}
\label{orthogonal}
\sum_{lm} U^k_{lml''m''}\,U^k_{lml'm'} = \delta_{l'l''} \delta_{m'm''}\,.
\end{equation}

\subsection*{$k^2=1$ modes}

For modes corresponding exactly to the curvature scale $q=0$, the 
normalization factor given by Eq.~(\ref{subNkl}) vanishes as $N_{1l} \propto 
q$, for all multipoles. In the absence of any physical reason why the 
amplitude of these modes should be zero, one can construct a homogeneous 
field using an alternative normalization factor 
\begin{equation}
\label{curvNkl}
\bar N_{1l} \equiv \lim_{q\to0} {N_{kl}\over q} = 
        \sqrt\frac2\pi \, {1 \over l!} \,,
\end{equation}
which leaves these modes finite. These modes are not normalizable, in the 
sense that the corresponding $\bar B_k =1/q^2$ in Eq.~(\ref{norm}) diverges 
as $q\to0$. However, orthogonality of the transformation matrix, 
Eq.~(\ref{orthogonal}), still holds in the analytic limit, which ensures the 
homogeneity of the field generated from the $k^2=1$ modes.

\subsection*{Super-curvature modes}

For super-curvature modes the wavenumber lies in the range $0<k^2<1$, so 
$q=\pm i\sqrt{1-k^2}$ is purely imaginary. The mode functions can be 
obtained from the sub-curvature modes by analytic continuation, giving
\begin{equation}
\label{Pikl_sup}
\tilde\Pi_{kl}(r) = -|q|^{-2} (\sinh r)^l \left(\frac{-1}{\sinh r} 
        \frac{d}{dr} \right)^{l+1} \cosh(|q|r) \,.
\end{equation}
These eigenfunctions extend beyond the curvature scale, and the resulting 
correlation function falls away exponentially only on scales greater than 
$1/k$~\cite{LW}. 

These mode functions are not normalizable, and neither are they
linearly independent of the sub-curvature modes in any finite region
(such as our observable universe). Nevertheless, they can be used to
construct a more general random field than is possible from the
sub-curvature modes alone~\cite{LW}. Because there is no orthogonality
relation as in Eq.~(\ref{norm}), one cannot fix the normalization of
these modes in the way we did for the sub-curvature modes. However, if
we use the analytic continuation of the normalization factor for the
sub-curvature modes
\begin{equation}
\label{supNkl}
N_{kl} = - \sqrt\frac2 \pi \, |q|^2 \, \prod^l_{s=0}(s^2-|q|^2)^{-1/2} \,,
\end{equation}
which is purely imaginary for all $l$, the transformation matrix $U$
will also be an analytic continuation. We know that Eq.~(\ref{unitary}) 
holds in the sub-curvature regime, and in fact it also holds in the 
super-curvature regime. At first sight it appears that we cannot appeal to 
the uniqueness of analytic continuation to demonstrate this, because the 
left hand side of Eq.~(\ref{unitary}) is not holomorphic. However, if real 
mode functions are used the unitarity relation becomes the orthogonality 
relation Eq.~(\ref{orthogonal}), which is preserved under the analytic 
continuation into the super-curvature regime. The transformation matrices 
$U$ remain real in this regime because {\it all} the $N_{kl}$ in
Eq.~(\ref{supNkl}) are purely imaginary, and so Eq.~(\ref{orthogonal})
implies unitarity and homogeneity.

\section{Positive-eigenvalue modes}

Modes with positive eigenvalues not only extend beyond the curvature
scale, but actually diverge as $r\to\infty$. However we have seen that
the super-curvature modes with $0<k^2<1$ can form a homogeneous random
scalar field despite being non-square integrable, so one might ask
whether a similar analytic continuation from the normalized
sub-curvature modes might also give a homogeneous random field for
$k^2<0$.

For $k^2<0$, the unnormalized radial functions given by Eq.~(\ref{Pikl_sup}) 
remain real; however the monopole normalization factor in Eq.~(\ref{supNkl}) 
is purely imaginary, while the dipole acquires an extra factor of $i$ and 
becomes real. Thus the normalized mode functions no longer have a unique 
phase. This means that the transformation matrices $U$ can no longer be
purely real, unless the modes of differing phase do not mix with one
another under a change of coordinates. Thus Eq.~(\ref{orthogonal}),
although it still holds, no longer guarantees unitarity,
Eq.~(\ref{unitary}).\footnote{We have not been able to prove that 
there is {\em no} normalization for which the transformation between modes 
with different phases becomes unitary, only that the analytic
continuation from sub-curvature modes does not give a unitary
transformation.} 

For instance, for $-3<k^2<0$, the $l\geq 1$ multipoles are all real but
the monopole is purely imaginary.  One might hope to build a homogeneous
random field from only the higher multipoles, but the orthogonality
relation, Eq.~(\ref{orthogonal}), involves a sum over all the
multipoles. Moreover a distribution with no monopole in one coordinate
system will in general acquire a monopole term after a shift of the
origin, which implies inhomogeneity.

More generally, for $1-(n+1)^2 < k^2 < 1-n^2$ (where $n\geq1$ is an
integer), the normalization given by Eq.~(\ref{supNkl}) has a phase 
$(l-1)\pi/2$ for $l<n$, while for all $l \geq n$ it has the same phase, 
$(n-1)\pi/2$. A shift of the origin mixes all the multipoles and any attempt 
to construct a random scalar field only from modes with the same phase fails 
to respect homogeneity.

\subsection*{Homogeneous tensor fields when $k^2=1-n^2$}

An interesting situation arises for the discrete set of modes when
$k^2=1-n^2$ with $n\geq1$ an integer.  These modes can be obtained
from the discrete {\em closed} universe modes with $k^2=n^2-1$ by
analytically continuing both the radial coordinate $r\to ir$ and the
wavenumber \mbox{$k\to ik$}~\cite{Har}. 

Note that Eqs.~(\ref{Pikl_sup}) and (\ref{supNkl}) give finite 
expressions for the first $n$ multipoles, while for the higher multipoles 
$N_{kl}\tilde{\Pi}_{kl}(r)={\cal O}(\epsilon)$, where 
$\epsilon^2=n^2-1+k^2$. An alternative is to define
\begin{eqnarray}
\label{GBNkl}
\bar N_{kl} & \equiv & \lim_{\epsilon\to0} \,\epsilon\, N_{kl} \,,\\
\label{GBPikl}
\bar \Pi_{kl} & \equiv & \lim_{\epsilon\to0} \,
        {\tilde\Pi_{kl}\over\epsilon^2} \,,
\end{eqnarray}
which gives finite expressions for the $l\geq n$ multipoles, while
with this normalization the lower multipoles diverge as $1/\epsilon$.
{}From Eqs.~(\ref{Pikl}) and~(\ref{GBPikl}), we have, for $l\geq n$,
\begin{equation}
\label{GBPiklexplicit}
\bar \Pi_{kl} = {1\over2n^3} \left(\sinh r\right)^l \left(
{-1\over\sinh r} {d\over dr}\right)^{l+1} \left( r\sinh nr \right) \,.
\end{equation}
Rescaling all the mode functions independently of $l$ and $m$ does not
change the transformation matrix $U$.

We show in Appendix~B that the matrix $U$ becomes block diagonal in
the limit $\epsilon\to0$, so that the orthogonality relation,
Eq.~(\ref{orthogonal}), holds separately for transformations between
the $l<n$ multipoles and for transformations between the $l\geq n$
multipoles.

In a closed universe it is well known that the $l<n$ multipoles form a
closed unitary group under coordinate transformations.  In an open
universe, while the $l<n$ modes still form a closed group, the mode
functions have alternating phases and so the transformation matrix is
not purely real. Thus the orthogonality relation no longer implies
unitary, and so one cannot form a homogeneous random scalar field.

By contrast, the $l\geq n$ multipoles all have the same phase in an
open universe (whereas in a closed universe they have alternating
phases). Hence the sub-matrix of $U$ connecting the higher multipoles
is real, and the transformation is unitary.  
One might think that it is possible to construct a homogeneous scalar
random field from the $l\geq n$ multipoles alone. However this is not
possible, because the lower multipoles can be regenerated by a
coordinate transformation. This is despite the fact that $U$ becomes
block diagonal, as the contribution from the diverging low multipoles
given in Eqs.~(\ref{GBNkl}) and (\ref{GBPikl}) remains finite in the
limit $k^2\to1-n^2$ even though the matrix elements approach zero. 

However if we act on the scalar field with an operator which kills the
lower multipoles, we obtain a homogeneous tensor random field even
though the underlying scalar field is inhomogeneous. We now discuss
two physical cases in which this does in fact occur.

\subsection*{$k^2=0$ modes}

As $k^2 \to 0$, the normalization factor
for the monopole in Eq.~(\ref{supNkl}) gives $N_{00} = i\sqrt{2/\pi}$,
while the higher multipoles diverge as $N_{0l} \propto i/k$. At the same
time, the unnormalized monopole $\tilde{\Pi}_{00} \to 1$, while the mode
functions given by Eq.~(\ref{Pikl_sup}) vanish for $l\geq1$ as
$\tilde\Pi_{0l} \propto k^2$. One can construct a finite field from the
multipoles $l\geq 1$ if we use the functions~\cite{LW,GZ}\footnote{Note
there is a typographical error in Eq.~(100) of Ref.~\cite{LW}.}
\begin{eqnarray}
\label{GZNkl}
\bar N_{0l} &\equiv& \lim_{k\to0} k N_{kl}\, ,\\
\bar \Pi_{0l} &\equiv& \lim_{k\to0} {\tilde\Pi_{kl}\over k^2}\, ,
\end{eqnarray}
which leaves these modes finite (although the monopole becomes
infinite with this normalization). Such a mode appears if
one considers the quantum fluctuations of a massless scalar field in
de Sitter space-time using an open universe coordinate
system~\cite{sasaki}. The anisotropy of the microwave background sky
due to curvature perturbations with $k^2=0$ in an open universe has
also been discussed recently~\cite{GZ}.

The monopole $N_{00} \Pi_{00}(r) Y_{00}(\theta,\phi)$ is a
constant when $k^2=0$. As a result {\em any} tensor field
constructed by covariant differentiation of the scalar field will be
homogeneous. The simplest example is the vector field
\begin{equation}
V^i \equiv \nabla^i f \,.
\end{equation}
Note that, in the notation of Ref.~\cite{Bardeen}, $V^i$ is
indistinguishable from an intrinsically `vector' quantity when $k^2 = 0$, as 
it is {\em solenoidal}, $\nabla_i V^i = 0$.

\subsection*{$k^2=-3$ modes}

Another interesting case of the $k^2=1-n^2$ modes discussed above is
$n=2$. In the limit $k^2 \to -3$, the monopole normalization is $N_{k0} = 
2i\sqrt{2/\pi}$ and that of the dipole is
$N_{k1} = 2\sqrt{2/3\pi}$, while the normalizations of the higher
multipoles diverge as $N_{kl}\propto1/\epsilon$, where $\epsilon^2=
3+k^2$.  On the other hand, the radial functions behave as
$\tilde\Pi_{k0} \to \cosh r$, $\tilde\Pi_{k1} \to - \sinh r$, and
$\tilde\Pi_{kl} \propto \epsilon^2$ for $l\geq 2$. We can use the
normalization given in Eqs.~(\ref{GBNkl}) and~(\ref{GBPikl}) to
render the $l\geq 2$ multipoles finite. 

To construct a homogeneous tensor random field in this case we need to kill 
both the monopole and the dipole. The simplest operator which does this is
\begin{equation}
\label{Tab}
T_{ij} \equiv \nabla_i \nabla_j - \frac{1}{3}
        \gamma_{ij}\nabla^2 \,,
\end{equation}
where $\gamma_{ij}$ is the spatial metric. The explicit form of the
tensor components for the metric given by Eq.~(\ref{lineelement}) are
given in Appendix~C. Acting on a scalar, this gives a traceless and
(for $k^2=-3$) transverse second-rank tensor. Because of the
transversality, there is no distinction between these `scalar' metric
perturbations and intrinsically `tensor' gravitational waves, in the
notation of Ref.~\cite{Bardeen}.  A physical example of this case is
the metric perturbation associated with quantum fluctuations of the
bubble wall in open inflation models~\cite{Hamazaki,bubble}.

The lowest non-vanishing modes are the quadrupole and octopole whose
radial dependence is given from Eq.~(\ref{GBPiklexplicit}) as
\begin{eqnarray}
\bar \Pi_{k2} &=& -\, {1\over8} \left( 2\cosh r -{3\cosh r\over\sinh^2r}
 +{3r \over \sinh^3r} \right) \, ,\\
\bar \Pi_{k3} &=& -\, {1\over8} \left( 2\sinh r - {5\over\sinh r}
 - {15\over\sinh^3r}
+\,{15r\cosh r \over \sinh^4r} \right) \, .\nonumber \\
\end{eqnarray}
Note that these higher multipoles of the scalar field diverge as
$\bar\Pi_{kl}\sim e^r$ as $r\to\infty$. The action of the tensor
operator renders some components of $T_{ij}$ finite at infinity, such
as $T_{rr}$, $T_{r\theta}$ and $T_{r\phi}$ which are of order
$e^{-r}$, but the remaining components of $T_{ij}$ still diverge as
$e^r$. Nonetheless, due to the form of the metric inverse
[$\gamma^{ij}={\rm diag}(1,1/\sinh^2r,1/\sin^2\theta\sinh^2r)$] this
is sufficient to leave scalar invariants finite at infinity, e.g.,
$T_{ij}T^{ij}\sim e^{-2r}$.

\section{Conclusions}

To summarize, the normalization defined by Eq.~(\ref{subNkl}) for
sub-curvature modes can be used to generate a homogeneous field
because of the orthogonality relation in Eq.~(\ref{norm}). We have
shown that it remains valid in the super-curvature regime, by virtue
of the fact that the radial mode functions for a given eigenvalue
$k^2$ have the same phase.

These normalized eigenfunctions can be written, for both sub- and
super-curvature modes, as
\begin{eqnarray}
\label{generalPikl}
\Pi_{kl}(r)&=& \left[\frac{\Gamma(l+1+iq)\Gamma(l+1-iq)}
        {\Gamma(iq)\Gamma(-iq)}\right]^{1/2} \times \nonumber\\
&& \hspace*{2cm} {P^{-l-1/2}_{iq-1/2} (\cosh r)\over\sqrt{\sinh r}}\,,
\end{eqnarray}
where $q^2=k^2-1$. For sub-curvature modes, these functions are
real, while for super-curvature modes they are purely imaginary.

If we were to multiply the mode functions by an \mbox{$l$-dependent}
{\em phase} in the sub-curvature regime, they would still be suitably
normalized there but the continuation of the phase factor to the
super-curvature regime would in general spoil the normalization of the
super-curvature modes. An example of this would be to replace the
first factor in Eq.~(\ref{generalPikl}) by $\Gamma(l+1+iq)/
\Gamma(iq)$. Both normalizations are equivalent in the sub-curvature
regime, but the latter is not suitable for analytic continuation to
the super-curvature regime, where it gives an incorrect normalization
and also fails to be symmetric under $q \leftrightarrow -q$.
Although Lyth and Woszczyna~\cite{LW} and Hamazaki et
al.~\cite{Hamazaki} both quoted this latter form for the
sub-curvature modes, they did not use it directly for analytic
continuation and in fact both these papers obtained the satisfactory
super-curvature mode normalization given by Eq.~(\ref{generalPikl}) above.

For $k^2\leq0$ there is no unique phase, and therefore one cannot use
the analytically continued mode functions to construct a homogeneous
{\em scalar} random field.  However, in the specific case $k^2=1-n^2$,
for integer $n$, the transformation between multipoles with $l\geq n$
is unitary, which allows a homogeneous {\em tensor} random field to be
constructed by acting on the scalar with a covariant differential operator,
provided that the operator kills the lower multipoles.  For $k^2=0$ we
have noted that any differential operator does this, since the
monopole is spatially constant. For $k^2=-3$ we have seen that the
traceless symmetric second-rank tensor does this.  A physical example
of the latter is the metric perturbation generated by bubble wall
fluctuations in open inflation models. The normalized eigenfunctions
for the higher multipoles, given in Eqs.~(\ref{GBNkl})
and~(\ref{GBPikl}), can be written as~\cite{bubble}
\begin{eqnarray}
\hat{\Pi}_{n l}(r) & = & \left[\frac{\Gamma(l+1+n)\Gamma(l+1-n)}{2}
        \right]^{1/2} \times \nonumber\\
&& \hspace*{2cm} {P^{-l-1/2}_{n-1/2} (\cosh r)\over\sqrt{\sinh r}}\,.
\end{eqnarray}

We recover the flat-space limit by taking $r \to 0$ while keeping $|q|r$ 
fixed. Only the sub-curvature modes survive in this limit, where they tend 
to the usual spherical Bessel functions~\cite{LW}. The super-curvature 
modes, which have $|q| < 1$, are not present in this limit, and the $l \geq 
n$ multipoles of the discrete modes with $k^2 = 1-n^2$ can be discarded in 
this limit since $n \to \infty$.

\section*{Acknowledgments}

J.G.B.~and D.W.~are supported by PPARC, and A.R.L.~by the Royal Society.
We thank Martin Bucher, Joanne Cohn, Misao Sasaki and Dominik 
Schwarz for useful discussions. J.G.B., A.R.L.~and D.W.~thank Stanford 
University for its hospitality during part of this work, with visits funded 
by NATO Collaborative Research Grant Ref.~CRG.950760. D.H.L.~thanks the 
University of California at Berkeley for its hospitality, and
D.W.~thanks the Astronomy Centre at the University of Sussex.

\appendix
\section{Mode Orthogonality and Transformation Unitarity}

Here we prove that the orthogonality of the sub-curvature modes
implies that the transformation induced by shifts in origin and
orientation of coordinates is unitary. An alternative proof was given 
in Ref.~\cite{LW}. This means that a random field generated from them is 
homogeneous.

\typeout{WARNING: LaTeX2e MAY RUN INCOMPATIBLY WITH RevTeX IN THE `array' 
ENVIRONMENT.} 
\typeout{IF YOU GET AN ERROR, JUST TYPE `q' OR `return' THROUGH IT,}
\typeout{AND ALL SHOULD BE WELL. Andrew Liddle}

Substituting the transformation 
\begin{equation}
Z_{klm}(r,\theta,\phi)=\sum_{l'm'} U^k_{lml'm'} 
        Z_{kl'm'}(r',\theta',\phi')\, ,
\end{equation}
into the orthogonality relation
\begin{equation}\label{norma}
\int dV  Z^*_{klm}({\bf x}) Z_{k'l'm'}({\bf x}) = B_k 
        \,\delta(|q|-|q'|) \, \delta_{ll'} \, \delta_{mm'} \, ,
\end{equation}
gives
\begin{eqnarray}
\int &dV&  Z^*_{klm}({\bf x}) Z_{k'l'm'}({\bf x}) \nonumber\\
        &=& \int dV \sum_{l''m''} (U^k_{lml''m''})^* 
        Z^*_{kl''m''}({\bf x}') \times \nonumber\\
 && \quad \quad \sum_{l'''m'''} U^k_{l'm'l'''m'''} 
        Z_{kl'''m'''}({\bf x}') \nonumber\\
        &=& \sum_{l''m''} (U^k_{lml''m''})^* U^k_{l'm'l''m''}
        B_k \,\delta(|q|-|q'|) \,  \, ,
\end{eqnarray}
where the last equality uses the orthogonality relation, Eq.~(\ref{norma}),
for the ${\bf x}'$ coordinates. Thus we have
\begin{equation}
\sum_{l''m''} (U^k_{lml''m''})^* U^k_{l'm'l''m''} = \delta_{ll'}\,
        \delta_{mm'} \, ,
\end{equation}
and hence the transformation is unitary.

Note that orthogonality also gives us an expression for the matrix 
$U^k_{lml'm'}$, namely
\begin{eqnarray}
\label{Uint}
\int &dV&  Z^*_{k'l'm'}({\bf x}') Z_{klm}({\bf x}) \nonumber\\
        &=& \int dV' Z^*_{k'l'm'}({\bf x}') \sum_{l''m''} U^k_{lml''m''} 
        Z_{kl''m''}({\bf x}')  \nonumber\\
        &=& U^k_{lml'm'} B_k \,\delta(|q|-|q'|) \,.
\end{eqnarray}

\section{Block diagonality of $U$ for \lowercase{$k^2=1-n^2$}}

In this Appendix we shall demonstrate that the matrix $U$ becomes
block diagonal in the limit $k^2\to1-n^2$ for integer $n\geq 1$. 
We split the transformation matrix $U$ into block matrix form
\begin{equation}
U=\left(\begin{array}{cc} A & B \\ C & D \end{array} \right)\,,
\end{equation}
where $A_{lml'm'}\equiv U^k_{lml'm'}$ for $l,l'<n$ and so on.
Then the orthogonality condition $UU^{\rm T} =I$ gives
\begin{eqnarray}
\label{AAT}
AA^{\rm T}+BB^{\rm T}&=& I\,,\\
AC^{\rm T}+BD^{\rm T}&=& 0\,,\\
\label{DBT}
CA^{\rm T}+DB^{\rm T}&=& 0\,,\\
\label{DDT}
CC^{\rm T}+DD^{\rm T}&=& I\,,
\end{eqnarray}
where $I$ is the identity matrix.

We adopt the normalization of the mode functions given in
Eqs.~(\ref{GBNkl}) and (\ref{GBPikl}), which leaves the $l\geq n$
multipoles finite but leads to the $l<n$ multipoles diverging as
$1/\epsilon$ where $k^2=1-n^2+\epsilon^2$. Consider a function
composed solely of modes with $k^2=1-n^2+\epsilon^2$
\begin{equation}
f_k({\bf x}) = \sum_{lm} f_{klm} \bar{N}_{kl} \bar\Pi_{kl}(r)
        Y_{lm}(\theta,\phi) \, .
\end{equation}
If this is to remain bounded at finite $r$ as $\epsilon \to 0$,
we must have $f_{klm}={\cal O}(\epsilon)$ for $l<n$ and $f_{klm}$
finite for $l\geq n$. Under a coordinate transformation
\begin{equation}
f_k({\bf x}) = \sum_{l'm'} f'_{kl'm'} \bar{N}_{kl'} \bar\Pi_{kl'}(r')
        Y_{l'm'}(\theta',\phi') \,,
\end{equation}
where $f'_{kl'm'}$ is given by Eq.~(\ref{fprime}), and we must likewise have
$f'_{kl'm'}={\cal O}(\epsilon)$ for $l'<n$. 

Thus the sub-matrix $C = {\cal O}(\epsilon)$. In the limit
$\epsilon\to0$, Eq.~(\ref{DDT}) becomes $DD^{\rm T}=I$ and
Eq.~(\ref{DBT}) becomes $DB^{\rm T}=0$. Because $D$ is orthogonal this
implies $B=0$, and hence the matrix $U$ is block diagonal. Finally, we
note from Eq.~(\ref{AAT}) that $AA^{\rm T}=I$.

\section{\lowercase{$k^2=-3$} tensor modes}

In this Appendix we give the actual components of the (symmetric)
operator $T_{ij}$ given in Eq.~(\ref{Tab}) for the
metric of Eq.~(\ref{lineelement}),
\begin{eqnarray}
T_{rr} & = & {\partial^2\over\partial r^2} - \frac{1}{3}\nabla^2 \,,\\
T_{\theta\theta} & = & {\partial^2\over\partial\theta^2} + \sinh r
\left( \cosh r {\partial\over\partial r} - \frac{1}{3} \sinh r
\nabla^2 \right) \,,\\
T_{\phi\phi} & = & {\partial^2\over\partial\phi^2} +
\sin\theta\cos\theta{\partial\over\partial\theta} \nonumber\\
&& \quad + \sin^2\theta \sinh r
\left( \cosh r {\partial\over\partial r} - \frac{1}{3} \sinh r
\nabla^2 \right) \,,\\
T_{r\theta} & = & \left( {\partial\over\partial r}
 - {\cosh r \over \sinh r} \right) {\partial\over\partial\theta} \,,\\
T_{r\phi} & = & \left( {\partial\over\partial r}
 - {\cosh r \over \sinh r} \right) {\partial\over\partial\phi} \,,\\
T_{\theta\phi} & = & \left( {\partial\over\partial\theta}
 - {\cos\theta\over\sin\theta} \right) {\partial\over\partial\phi} \,.
\end{eqnarray}

The unnormalized $k^2=-3$ monopole and dipole modes are given, using
Eq.~(\ref{Pikl_sup}), by
\begin{eqnarray}
\tilde\Pi_{k0}(r)Y_{00}(\theta,\phi)
& = & \sqrt{1\over4\pi} \cosh r \,, \\
\tilde\Pi_{k1}(r)Y_{10}(\theta,\phi)
& = & - \sqrt{3\over4\pi} \sinh r \cos\theta \,, \\
\tilde\Pi_{k1}(r)Y_{11}(\theta,\phi)
& = & \sqrt{3\over4\pi} \sinh r \sin\theta \cos\phi \,,\\
\tilde\Pi_{k1}(r)Y_{1(-1)}(\theta,\phi)
& = & \sqrt{3\over4\pi} \sinh r \sin\theta \sin\phi \,.
\end{eqnarray}
It is then straightforward to verify that each component of $T_{ij}$
vanishes everywhere when applied to the monopole and dipole modes.
Thus the action of the operator in Eq.~(\ref{Tab}) on the $k^2=-3$
modes can indeed form a homogeneous tensor field from the $l\geq2$
multipoles of the scalar field whose normalized radial functions are
given by Eqs.~(\ref{GBNkl}) and~(\ref{GBPikl}).


\end{document}